\documentstyle[twoside,12pt]{article}

\catcode`\@=11
\long\def\@makefntext#1{ 
\protect\noindent \hbox to 3.2pt {\hskip-.9pt
$^{{\eightrm\@thefnmark}}$\hfil}#1\hfill} 

\def\thefootnote{\fnsymbol{footnote}}
 \def\@makefnmark{\hbox to 0pt{$^{\@thefnmark}$\hss}}  

\def\ps@myheadings{\let\@mkboth\@gobbletwo
\def\@oddhead{\hbox{} 
\rightmark\hfil\eightrm\thepage}
\def\@oddfoot{}\def\@evenhead{\eightrm\thepage\hfil 
\leftmark\hbox{}}\def\@evenfoot{}
\def\sectionmark##1{}\def\subsectionmark##1{}}

\oddsidemargin=\evensidemargin
\renewcommand{\thefootnote}{\fnsymbol{footnote}}

\newcounter{sectionc}\newcounter{subsectionc}\newcounter{subsubsectionc}
\renewcommand{\section}[1] {\vspace{12pt}\addtocounter{sectionc}{1}
\setcounter{subsectionc}{0}\setcounter{subsubsectionc}{0}\noindent
	{\bf\thesectionc. #1}\par\vspace{5pt}}
\renewcommand{\subsection}[1] {\vspace{12pt}\addtocounter{subsectionc}{1}
	\setcounter{subsubsectionc}{0}\noindent
	{\bf\thesectionc.\thesubsectionc. {\kern1pt \bf\it #1}}\par\vspace{5pt}}
\renewcommand{\subsubsection}[1] {\vspace{12pt}\addtocounter{subsubsectionc}{1}
	\noindent{\thesectionc.\thesubsectionc.\thesubsubsectionc.
	{\kern1pt \it #1}}\par\vspace{5pt}}
\newcommand{\nonumsection}[1] {\vspace{12pt}\noindent{\bf #1}
	\par\vspace{5pt}}

\newcommand{\textlineskip}{\baselineskip=14pt}

\def\eightcirc{
\begin{picture}(0,0)
\put(4.4,1.8){\circle{6.5}}
\end{picture}}
\def\eightcopyright{\eightcirc\kern2.7pt\hbox{\eightrm c}}

\def\abstracts#1#2#3{{
	\centering{\begin{minipage}{5in}\baselineskip=12pt\tenrm
	\centerline{ABSTRACT}
	\parindent=0pt #1\par
	\parindent=15pt #2\par
	\parindent=15pt #3
	\end{minipage} }\par}}

\renewenvironment{thebibliography}[1]			
	{
	 \begin{list}{\arabic{enumi}.}			
	{\usecounter{enumi}\setlength{\parsep}{0pt}
	 \setlength{\leftmargin 17pt}{\rightmargin 0pt}	
	 \setlength{\itemsep}{0pt} \settowidth		
	{\labelwidth}{#1.}\sloppy}}{\end{list}}	

\newcounter{itemlistc}
\newcounter{romanlistc}
\newcounter{alphlistc}
\newcounter{arabiclistc}

\newcounter{tempfigtabc}			

\def\pmb#1{\setbox0=\hbox{#1}
	\kern-.025em\copy0\kern-\wd0
	\kern.05em\copy0\kern-\wd0
	\kern-.025em\raise.0433em\box0}

\def\fnt#1#2{\footnotetext{\kern-.3em
	{$^{\mbox{\scriptsize #1}}$}{#2}}}

\def\fpage#1{\begingroup
\voffset=.3in
\thispagestyle{empty}\begin{table}[b]\centerline{\footnotesize #1}
	\end{table}\endgroup}

 1
 1
 1

\font\eightrm=cmr8

\def\qed{\hbox{${\vcenter{\vbox{                          
   \hrule height 0.4pt\hbox{\vrule width 0.4pt height 6pt
   \kern5pt\vrule width 0.4pt}\hrule height 0.4pt}}}$}}

\def\I{{\cal I}}
\def\del{\partial}
\def\Tr{{\rm Tr}}
\def\Lsu{{\cal L_{\mbox{\tiny SU(2)}}}}
\def\bea{\begin{eqnarray}}
\def\be{\begin{eqnarray}}
\def\ee{\end{eqnarray}}
\def\eea{\end{eqnarray}}
\def\ben{\begin{enumerate}}
\def\een{\end{enumerate}}
\def\bitem{\begin{itemize}}
\def\eitem{\end{itemize}}

\def\eg{{\it e.g.}}
\def\L{{\cal L}}

\def\bR{{\bf R}}
\def\bT{{\bf T}}
\def\bA{{\bf A}}

\def\hatR{\hat{R}}
\def\bJ{{\bf J}}
\begin{document}
\normalsize\textlineskip
{\thispagestyle{empty}
\setcounter{page}{1}

\renewcommand{\thefootnote}{\fnsymbol{footnote}} 
\def\bsc{{\sc a\kern-6.4pt\sc a\kern-6.4pt\sc a}}
\def\bflatex{\bf L\kern-.30em\raise.3ex\hbox{\bsc}\kern-.14em
T\kern-.1667em\lower.7ex\hbox{E}\kern-.125em X}

\fpage{1}
\centerline{\bf HEAVY-QUARK BARYONS AS SKYRMIONS \footnote{Invited
talk given at {\it the International Workshop on Baryons as Skyrme Solitons},
27-30 September 1992, Siegen, Germany}}
\vspace{0.37truein}
\centerline{\footnotesize Mannque RHO}
\vspace*{0.015truein}
\centerline{\footnotesize\it Service de Physique Th\'{e}orique, C.E. Saclay}
\baselineskip=12pt
\centerline{\footnotesize\it 91191 Gif-sur-Yvette, France}

\vspace*{0.21truein}
\abstracts{ I discuss recent development on the description of
heavy-quark (such as charmed and bottom) baryons as one or more
heavy mesons ``wrapped" by a skyrmion. Amazingly enough, such a description
naturally arises when light-quark chiral symmetry and heavy-quark
spin symmetry are incorporated in an effective Lagrangian. I interpret
the resulting spectrum in terms of nonabelian induced (Berry) potentials
in analogy to diatomic molecular systems.
}{}{}

\vspace*{-3pt}\textlineskip
\section{Introduction}

There is a wide-spread feeling that the skyrmion description of the
baryons is a highly predictive and coherent one, ranging from the
structure of the nucleon and $\Delta$ to nuclear forces. Indeed many
of the no-go arguments raised against the skyrmion picture are disappearing
rapidly.
As we heard in this meeting, we no longer seem to have any problem with
S- and P-wave $\pi$N scatterings$^{1,2}$; yesterday's too large energy of the
ground state
baryon seems no longer too large once Casimir contributions of $O(N_c^0)$
are taken into account$^{1,3}$; the missing attraction at medium range
in NN potentials that binds nuclei
is no longer missing$^{4}$; the soliton instability problem
is just a red herring; whether explicit quark degrees of freedom should be
present or not is becoming a non-issue. At this moment, there seem
to exist no really
serious arguments against the notion that the nucleon is a skyrmion in
and outside of nuclear medium. Whatever deviation from nature there may be,
may simply be a matter of poorly understanding the intricacy of the
skyrmion, not the deficiency of the model itself. As far as I know, it is
the only model available
which can address simultaneously one-nucleon {\it and} many-nucleon
problems with equal ease. It is of course not derived from QCD but it
is consistent with it.

In this talk, I will argue that even when one has very heavy quarks in the
baryons, the skyrmion picture still makes sense and in fact, it
comes out strikingly close to the quark-model description which we
expect gets better as the quark mass becomes heavier. This may sound amazing
or, to some, unbelievable. In fact, when Riska, Scoccola and I submitted a
paper
a few years ago showing that the skyrmion model worked just as well
in charmed and bottom baryon sectors, a referee promptly rejected it. It
appeared in a different journal$^5$ only after a long delay because of the
referee's repeated
refusal to accept our arguments. As I shall describe below, our claim is
vindicated by the recent development.
\textheight=8.5truein
\setcounter{footnote}{0}
\renewcommand{\thefootnote}{\alph{footnote}}

\section{Diatomic Molecules}

To better clarify the basic idea involved in the workings of the soliton
model in complex strong interactions, let me start with a quantum mechanical
problem of a diatomic molecule. This quantum mechanical problem can be put in
a context that closely mimics the baryon excitation we are interested in.
It shows a generic setting in which nonabelian gauge potentials are {\it
induced} by interactions.

Consider the dynamics of a diatomic molecule where two atoms are separated
by the internuclear separation denoted by the vector $\vec{R}$.
We shall restrict ourselves to the relative motion only. Because of the
symmetry of the diatom, rotational symmetry of the electrons is broken.
The electronic state is characterized by the eigenvalue $\Omega$
of the operator $\hatR\cdot {\bJ}_{el}$. In my discussion, I will confine
myself to the triplet of the states $\Omega=0, \pm 1$. This system was
recently studied by Zygelman$^6$ and reanalyzed by Lee and myself$^7$
to gain useful insight into its generic structure.

This system can be roughly categorized into two according to the size of
$R$. For small $R$, the potential curve for the $\Omega=0$ state
(called $\Sigma$) which is
higher lying does not overlap with that of the degenerate doublet $\Omega=
\pm 1$ (called $\pi$). Thus one can focus on the doublet, ignoring the singlet.
For large $R$, however, the potential curves start overlapping, the complete
overlap occurring at $R=\infty$. Then the triplets become degenerate,
with the {\it restoration of the electronic rotational symmetry.}
We will exploit this feature later on in discussing heavy-quark symmetry.

Let me now describe how the above structure can be understood.$^7$
We will consider the molecular excitation described by the dynamical variables
${\bR} (t)$ which could be vibration or rotation as {\it slow}
compared with the electronic excitation which we consider to be {\it fast}.
We wish to integrate out the fast degree of freedom and express the whole
system on the coordinate ${\bR} (t)$ of the slow degree of freedom.
The resulting system can be described by the following Lagrangian
\bea
{\cal L}=\frac{1}{2}\mu\dot{\vec{R}}^2 + i\theta^{\dagger}_a(\frac{\partial}
{\partial t}-i\vec{A}^{\alpha}T^{\alpha}_{ab}\cdot\dot{\vec{R}})\theta_b
\label{dimol}
\eea
where $\mu$ is the reduced mass,
$\theta_a$ is a Grassmann variable labeled by $a$, $\vec{A}^\alpha$ is
the Berry potential inherited from integrating out the fast degree of
freedom and $T^\alpha$ is the matrix representation of the vector space
in which the Berry potential lives. In our case, we have a triplet of
states and ${\bA}\cdot{\bT}\in SU(2)$. I have kept only the essential
pieces in the Lagrangian (\ref{dimol}), leaving out some ``junks",
to make the argument as simple as possible. The ``junks" do not change
the main thrust of my argument.
Let me also mention that the Grassmannian variables for
each $a$ are used as a trick to avoid writing the Lagrangian in matrix
form. There is nothing deep in it at least for our purpose.
Quantizing (\ref{dimol}) in a standard way, one gets the Hamiltonian
\bea
H=\frac{1}{2\mu} (\vec{p} - \vec{{\bf A}})^2 \label{dimolH}.
\eea
Since there is gauge invariance in the theory
as one can see from (\ref{dimol}),
we are allowed to make a gauge transformation on (\ref{dimolH})
and obtain the following Hamiltonian $^7$
\bea
H = -\frac{1}{2\mu R^2}\frac{\partial}{\partial R}R^2\frac{\partial}
{\partial R}
+ \frac{1}{2\mu R^2}(\vec{L}_o + (1-\kappa)\vec{I})^2
- \frac{1}{2\mu R^2}(1-\kappa)^2(\vec{I}\cdot\hat{R})^2\label{theH}
\eea
and the corresponding gauge field \footnote{We should note that the magnetic
field is of the nonabelian monopole type of 't Hooft and Polyakov with
however a charge which is {\it not} quantized. This is a generic feature
of induced gauge fields we encounter in various systems.}
$$\vec{{\bf A}}' = (1-\kappa)\frac{\hat{R} \times \vec{I}}{R^2},$$
and the magnetic field
$$\vec{{\bf B}}'=-(1-\kappa^2)\frac{\hat{R}(\hat{R}\cdot{\bf I})}
{R^2}$$
where $\kappa$ indicates the extent to which electronic
rotational symmetry is present, with $\kappa=1$ indicating full symmetry
and $\kappa=0$ a complete absence of symmetry.
In (\ref{theH}), $\vec{L}_o$ is the angular momentum lodged in the dumb-bell
and $\vec{I}$ is the angular momentum contributed by the Berry potential.
Neither $\vec{L}_o$ nor $\vec{I}$ separately commutes with the Hamiltonian.
What commutes is the the total angular momentum $\vec{L} =
\vec{L}_o + \vec{I}$.

We are now ready to analyze what happens
in the two extreme cases of $\kappa = 0$ which results when $R\rightarrow 0$
and $\kappa=1$ which results when $R\rightarrow \infty$.
For $\kappa = 1$, the degenerate $\Sigma$
and $\pi$ states form a representation of the rotation group and hence the
Berry potential (and its field tensor) vanishes or becomes a pure gauge.
The spectrum is then independent of the angular momentum carried by the
electronic system. {\it This just means that rotational symmetry is restored
in the electronic sector.}
For $\kappa=0$, the
$\Sigma$ and $\pi$ states are completely decoupled and only the
$U(1)$ monopole field can be developed on the $\pi$
states. $\kappa$ goes to zero and the Hamiltonian can be written as
\be
H = -\frac{1}{2\mu R^2}\frac{\partial}{\partial R}R^2\frac{\partial}
{\partial R}
+ \frac{1}{2\mu R^2}(\vec{L}\cdot\vec{L} - 1)\label{hzgu1}
\ee
which is a generic form for a system coupled to an $U(1)$ monopole field.
In this case, the ``magnetic charge" is quantized to $\pm 1$ or twice
the basic Dirac monopole charge $\pm \frac{1}{2}$.
A truly nonabelian Berry potential with non-quantized charge can be obtained
only for $\kappa$ which is not equal to zero or one.

\section{Spectrum of Heavy Baryons}

I will now turn to the structure of heavy baryons described as skyrmions
and summarize the recent development. What I will present below
is based on work done
recently in collaboration with Min and his coworkers in Seoul.$^8$
Instead of making detailed analysis to compare with experiments or
with quark models, let me start with a simplified Lagrangian consistent
with hidden gauge symmetry (HGS).$^9$ I will consider two light flavors $q=u,d$
and a third flavor $Q$ which will be taken to be heavy later on. For
the moment I will consider $u, d, Q$ on the same footing and
write a Lagrangian built from chiral symmetry. Obviously when the
mass of $Q$, $m_Q$, becomes large compared with the chiral scale
$\Lambda_\chi\sim 1$ Gev, it would make no sense to talk about chiral
symmetry associated with the quark $Q$ but let me blindly start with
an $SU(3)$ chiral Lagrangian anyway and take $m_Q$ become large.
I will write the Lagrangian  as
the sum of the $SU(2)$ Skyrme Lagrangian of the $(u,d)$ sector, $\Lsu$,
the HGS Lagrangian
without (with) the $\omega$ meson coupling, $\L_\Phi^{\mbox{\tiny HGS}}$
$(\L_\omega^{\mbox{\tiny HGS}})$, and the ``anomalous parity"
 Lagrangian, ${\L}_{an}$;
\be
\L^{\mbox{\tiny HGS}} &=& \Lsu + \L_\Phi^{\mbox{\tiny HGS}}
               + \L_\omega^{\mbox{\tiny HGS}} + \L_{an},
\nonumber \\
\Lsu &=& \frac{F_\pi^2}{16}
\Tr \left[ \del_\mu \Sigma \del^\mu \Sigma^\dagger
\right] + \frac{1}{32e^2} \Tr \left[ \Sigma^\dagger \del_\mu \Sigma ,
\Sigma^\dagger \del_\nu \Sigma \right]^2 ,   \nonumber \\
\L_\Phi^{\mbox{\tiny HGS}} &=&
\left(D_\mu \Phi \right)^\dagger \left( D^\mu \Phi \right)
- m_\Phi^2 \Phi^\dagger \Phi \nonumber \\
&& - \frac{1}{2} \Phi^{*\dagger}_{\mu\nu} \Phi^{*\mu\nu} +
m_{\Phi^*}^2 \left[ \Phi^{*\dagger}_\mu + \frac{2i}{F_\pi g_{\Phi^*}}
\Phi^\dagger A_\mu \right] \left[ \Phi^{*\mu} + \frac{2i}{F_\pi g_{\Phi^*}}
A^\mu \Phi \right]       \nonumber \\
\L_\omega^{\mbox{\tiny HGS}} &=& -\frac{iN_c}{2F_\pi^2} B_\mu
\left[ \left(  \Phi^\dagger D^\mu \Phi
       - (D^\mu \Phi)^\dagger \Phi \right)
 - \left( \Phi_\nu^{*\dagger} D^\mu \Phi^{*\nu}
- (D^\mu \Phi^*_\nu)^\dagger \Phi^{*\nu} \right) \right] \nonumber \\
{\L}_{an} &=&-  \frac{iN_c}{F_\pi^2} B_\mu
\left(  \Phi^\dagger D^\mu \Phi
       - (D^\mu \Phi)^\dagger \Phi \right)+\delta {\L}_{an}
\label{Lag:HGS}
\ee
where
\be
D_\mu &=& \del_\mu + V_\mu, \hskip 1cm \Sigma = \xi \cdot \xi, \nonumber \\
\left( \begin{array}{c} V_\mu \\ A_\mu \end{array} \right) &=& \frac{1}{2}
\left( \xi^\dagger \del_\mu \xi \pm \xi \del_\mu \xi^\dagger \right),
\nonumber \\
B_\mu &=& \frac{1}{24\pi^2} \epsilon_{\mu\nu\alpha\beta} \Tr \left\{
\Sigma^\dagger \del^\nu \Sigma \Sigma^\dagger \del^\alpha \Sigma
\Sigma^\dagger
\del^\beta \Sigma \right\}, \nonumber \\
\Phi^*_{\mu\nu} &=& \del_\mu \Phi^*_\nu - \del_\nu \Phi^*_\mu
+ V_\mu \Phi_\nu^* - V_\nu \Phi_\mu^*,
\ee
with $\epsilon_{0123}=+1$. Here, $\Sigma$ is the $SU(2)$ chiral field,
$\Phi$ and $\Phi^*_\mu$  are, respectively, the pseudoscalar and vector
meson doublets of the form $Q\bar{q}$, $F_\pi$ represents
the pion decay constant and
$g_{\Phi^*}$ is the $\Phi^*$ ``gauge" coupling to matter fields.
For instance, if we take the kaons to be heavy mesons,
$\Phi^\dagger = (K^-, \overline{K}^0)$,
$\Phi_\mu^{*\dagger} = (K_\mu^{*-}, \overline{K}_\mu^{*0})$.
This Lagrangian is obtained from a hidden gauge symmetric Lagrangian
$^{10}$ by integrating out the $\omega$ and $\rho$ meson fields and then
taking the limit
$m_\Phi= m_{\Phi^*} \to \infty$, neglecting the terms that
vanish as $m_\Phi^{-1}$ and $m_{\Phi^*}^{-1}$ or faster. For the purpose of
comparing with the heavy-quark limit $^{11}$ which we will refer to as
Isgur-Wise (IW) symmetric limit, it is necessary to keep the
vector mesons explicitly instead of integrating them out as done in
Scoccola et al.$^{10}$ The reason for this will become clear later on.

We need to explain a bit what ${\L}_{an}$ is in the context of the heavy-meson
limit that we are interested in. The first term is what one obtains from the
topological Wess-Zumino term written down by Witten $^{12}$ when expanded
\`{a} la Callan-Klebanov (CK).$^{13}$ This is intrinsically tied to anomalies
in
effective theory. Later, as the heavy quark mass increases, this term
will disappear. However the second term, which is intrinsic-parity odd
as the Wess-Zumino term is and involves the vectors $P^*$'s, need not
vanish in the heavy-quark limit. We expect it to modify the constants
of the main term responsible for the binding of the mesons $\Phi$ and
$\Phi^*$ to a soliton. They are fixed at low energy by low-energy theorems
and in the heavy-quark limit by heavy-quark symmetry. {\it A priori},
there is no reason why they should be related.

To see what remains of the Lagrangian (\ref{Lag:HGS}) in the
heavy-quark limit, we make the meson-field redefinition,$^{14}$
(taking $m = m_\Phi = m_{\Phi^*}\rightarrow \infty$)
\be
\Phi^{*\dagger}_\mu &=& e^{-im v \cdot x} P_\mu^* / \sqrt{m}, \nonumber \\
\Phi^\dagger &=& e^{-im v \cdot x} P / \sqrt{m},
\label{field}
\ee
so that the fields $P_\mu^*$ and $P$ are independent of the
meson mass and obtain
\be
\L_\Phi^{\mbox{\tiny HGS}} &=&
- i P v \cdot \stackrel{\leftrightarrow}{D} P^\dagger
+i P^*_\mu v \cdot \stackrel{\leftrightarrow}{D} P^{*\mu\dagger}
+ i \sqrt{2} \left( P A^\mu P^{*\dagger}_\mu
+ P^*_\mu A^\mu P^\dagger \right),
\label{HGS-lag} \\
\L_\omega^{\mbox{\tiny HGS}} &=&
\frac{N_c}{F_\pi^2} B_\mu
\left( P v^\mu P^\dagger - P_\nu^{*} v^\mu P^{*\nu\dagger}
\right) \label{LWZT} \\
{\L}_{an} &=&  \delta {\L}_{an}\label{LWZ}
\ee
where
\be
(D_\mu P)^\dagger &=& ( \del_\mu + V_\mu ) P^\dagger, \nonumber \\
A \stackrel{\leftrightarrow}{D} B^\dagger
&=& A (DB)^\dagger - (DA) B^\dagger.
\ee
We have not written out the term $\delta {\L}_{an}$ since while the
coefficients are known phenomenologically in the light-quark sector,
they are not known in the regime we are concerned with. We expect that
it will include terms of the form
\be
\frac{iN_c}{F_\pi \pi^2} \epsilon^{\mu\nu\alpha\beta} v_\mu
\left(a P A_\nu A_\alpha P_\beta^{*\dagger} -
b P_\beta^* A_\alpha A_\nu P^\dagger
\right)
\ee
with $a$ and $b$ unknown constants.
Note that since in $^9$ we started with an apparently $SU(3)$ symmetric
Lagrangian (apart from the meson mass term)
with the flavor $Q$ put on the same footing as the light quarks,
Eq. (\ref{LWZT}) results from the $\omega$-meson
coupling terms and hence the constant $N_c/F_\pi^2$ is fixed.
In the heavy-quark limit, the ``primordial" Wess-Zumino term is absent.
However, the rest will
remain to modify effectively
the coefficient of Eq.(\ref{LWZT}) which came
from the $\omega$-meson coupling with the heavy mesons $P$ and $P^*$.
That such a term must be present can be seen by bosonizing light and
heavy quarks from QCD.$^{15}$

If the $P$ and $P^*$ are degenerate,
the intrinsic-parity odd Lagrangian in which the Wess-Zumino term figures
in the HGS Lagrangian$^8$ contains a term that survives
in the heavy-meson mass limit
\be
c_4 \L_{(4)} = - c_4 2i g_{\Phi^*}^2
\epsilon^{\mu\nu\alpha\beta} v_\mu P_\nu^{*} A_\alpha P^{*\dagger}_\beta,
\label{four-d}
\ee
with $c_4$ the coefficient of $\L_{(4)}$.$^8$ This is effectively
a four-derivative term that belongs to the same intrinsic-parity class
as $\delta {\L}_{an}$ discussed above. The coefficient $c_4$ is fixed
in the light-quark sector to $c_4=iN_c/16\pi^2$ from the decay
$\omega\rightarrow \rho \pi$. We will see what the coefficient is
in the heavy-quark sector.

Our Lagrangian (\ref{HGS-lag})--(\ref{four-d}) can now be
compared with the one implied by IW symmetry $^{16,13}$
\be
\L_\Phi^{\mbox{\tiny JMW}} &=& -i \Tr \overline{H}_a v^\mu \del_\mu H_a
 + i \Tr \overline{H}_a H_b v^\mu \left(V_\mu \right)_{ba} \nonumber \\
& & + ig \Tr \overline{H}_a H_b \gamma^\mu \gamma_5
\left( A_\mu \right)_{ba} + \cdots \ \ ,  \label{LH}
\ee
with the heavy meson field $H_a$ (where $a$ labels the light-quark flavor)
defined as
\be
H = \frac{ ( 1 + \not\!{v})}{2} \left[ {P^*}_{\mu} \gamma^\mu -
P \gamma^5 \right]. \label{Ha}
\ee
In terms of $P$ and $P^*$, (\ref{LH}) reads
\be
\L_\Phi^{\mbox{\tiny JMW}}
&=& - i P v \cdot \stackrel{\leftrightarrow}{D} P^\dagger
+ i P^{*}_\mu v \cdot \stackrel{\leftrightarrow}{D} P^{*\mu\dagger}
\nonumber \\
&& + 2ig \left\{ P^{*}_\mu A^\mu P^\dagger + P A^\mu P_\mu^{*\dagger} \right\}
+ 2g \epsilon^{\lambda\mu\nu\kappa} v_\lambda P^{*}_\mu A_\nu
P^{*\dagger}_\kappa.
\label{Lag:Yan}
\ee
We see that the HGS Lagrangian (\ref{HGS-lag}) with
(\ref{four-d}) is identical -- except for the term (\ref{LWZT}) --
to the IW symmetric Lagrangian (\ref{Lag:Yan})
if we identify $g = 1 / \sqrt{2}$
and $c_4 g_{\Phi^*}^2=ig$. Surprisingly $g=1/\sqrt{2}$ is rather close to
the quark-model value $g\sim 0.75$ and also to the recent CLEO
collaboration data $^{17}$ $g\approx 0.6$. Furthermore low-energy
theorem~$^8$ gives $c_4 g_{\Phi^*}^2 \approx i0.65\approx ig$.
In (\ref{Lag:Yan}), the term of the form
(\ref{LWZT}) is missing for the simple reason that $B_\mu$ involves
three derivatives, so higher order in derivative expansion. Ignoring it
in  pion dynamics may be justified but it is not in soliton dynamics.
In fact we will see later that it is the most important contribution in our
way of describing heavy baryons as it is in the Callan-Klebanov scheme.$^{13}$
Since we do not know its coefficient in the IW limit, we will take it
in the form
\be
\L^{\mbox{\tiny HGS}}_\omega &=&  \alpha B_\mu j^\mu,\label{LWZP} \\
j^\mu &=& \Tr \left( \overline H v^\mu H \right).
\ee
with $\alpha$ an unknown constant. The Lagrangian (\ref{LWZP})
obviously satisfies both chiral symmetry and IW symmetry.
Such a term arises in an approximate bosonization of QCD, through the coupling
of $H$ to the $\omega$ meson.$^{15}$
In (\ref{LWZP}), $j^\mu$ is the $U(1)$ current of the
Lagrangian $\L_\Phi^{\mbox{\tiny JMW}}$ corresponding to the heavy-quark
flavor which is conserved in our case.
Although as mentioned above, the coupling constant $\alpha$ cannot be
determined
by chiral and Isgur-Wise symmetries alone, we will analyze the structure
of heavy baryons in units of $-N_c / 2F_\pi^2$, {\it i.e.}, the coefficient of
${\L}_\omega^{\mbox{\tiny HGS}}$ in Eq.(\ref{LWZT}).
We will normalize the meson field as
\be
\int \! d^3 r j^0 = -2  \int \! d^3 r
\left( P P^\dagger + P_i^{*} P^{*\dagger}_i \right)
= -1 \label{normalization}
\ee
and work in the rest frame of the heavy meson, $v_\mu = (1,0,0,0)$.
Note that $P_0^* = 0$ since $v \cdot P^* = 0$.

The Lagrangian correct to order $O(m_\Phi^0 \cdot N_c^0)$ is given by
\be
L_B &=&- {\cal M}_{sol}-m_\Phi + \int \! d^3 r( {\cal L}_P+ \L_W ) ,
\nonumber \\
-{\cal L}_P &=&  2gi \left\{ P^{*i} A^i P^\dagger
+ P A^i P^{*i\dagger}
- i \epsilon^{0ijk} P^{*i} A^j P^{*k\dagger} \right\} \nonumber \\
- \L_W &=&  2 \alpha B_0 \left( P P^\dagger + P^{*}_i P^{*\dagger}_i
\right).
\label{Ham}
\ee
One can readily see that ${\cal L}_P$ and ${\cal L}_W$ are invariant with
respect to the global
rotation $S\in SU(2)_V$ in the light flavor space ({\it i.e.}, the isospin
space) provided that the fields transform
\be
P(x) &=&  \phi (x) S^\dagger,  \nonumber \\
P_i^* (x) &=&  \phi_i^* (x)  S^\dagger, \nonumber \\
\xi (x) &=& S \xi_0 (\vec x) S^\dagger,
\label{trans}
\ee
with $x = (t,\vec x)$ and $\xi_0 (\vec x) = \exp ( i \vec \tau \cdot \hat r
F(r) / 2)$ in the hedgehog configuration.
Following the standard procedure for collective
quantization, we
elevate $S$ to a dynamical variable by endowing it with the time dependence
$S(t) = a_0 (t) + \vec a (t) \cdot \vec \tau$. As defined, the
fields $\phi (x)$ and $\phi_i^* (x)$ are fields living in the rotating
frame

The equations of motion for $\phi (x)$ and $\phi^* (x)$ gotten from
the Lagrangian valid at $O(m_\Phi^0 \cdot N_c^0)$ imply that
\be
| \phi (x) |^2  &\propto& \delta ^3 (\vec{x}), \nonumber\\
|\phi^*_i (x)|^2  &\propto& \delta ^3 (\vec{x}).\nonumber
\ee
Given these solutions, the energy shift
coming from $-({\cal L}_P + \L_W)$ of (\ref{Ham}) is
\be
E_I &=&- \int ({\cal L}_P + \L_W) \! d^3 r \nonumber \\
&=& - \frac{1}{2\pi^2} \alpha  \left\{ F'(0) \right\}^3
\label{energy}
\ee
{\it with the contribution of $\L_P$ vanishing identically.}
This differs from the
result of the recent work by Manohar and his collaborators$^{18}$ who
get the entire action from $\L_P$ whose contribution is non-vanishing since
the meson $H$ is not rotated in their scheme in contrast to our scheme
(\ref{trans}). I will discuss the difference of these two approaches later.

\subsection{Spectrum in IW limit}

We take $g = 1/\sqrt{2}
\simeq 0.7$, $F^\prime (0)\approx
-0.89 $ GeV from the literature and the experimental value of
$F_\pi = 186$ MeV and $e=4.75$, with which
we find the $\alpha$ value in the b-quark sector should be
\be
\alpha \approx - \frac{1}{2.8} \left( \frac{N_c}{2 F_\pi^2} \right)
\ee
to reproduce $M_{\Lambda_b} - M_{\mbox{\tiny N}} = 4.65$ GeV,  the
predicted value of the quark model. This corresponds to the binding energy of
\be
E_I\approx -0.55 {\mbox{ GeV}}.
\ee

Next we consider the effects of $O(m_\Phi^0 \cdot N_c^{-1})$ term in the
Lagrangian. For this we define
\be
S^\dagger \dot S = i \vec \tau
\cdot \vec \Omega
\ee
and write the corresponding Lagrangian to $O(m_\Phi^0 \cdot N_c^{-1})$
\be
L_{(-1)} = \int \! d^3 r \L_{(-1)} = 2 \I \Omega^2 - 2 \vec \Omega \cdot \vec
Q,
\label{canonical}
\ee
where
\be
\vec Q &=& - \int \! d^3 r \left( \phi \vec n (\vec r) \phi^\dagger +
\phi_i^{*} \vec n (\vec r) \phi_i^{*\dagger} \right), \nonumber \\
\vec n &=& \frac{1}{2} \left( \xi_0^\dagger \vec \tau \xi_0 + \xi_0 \vec \tau
\xi_0^\dagger \right) \nonumber \\
&=& \cos F(r) \vec \tau - (\cos F(r) - 1 ) \hat r ( \vec \tau \cdot \hat r ),
\ee
and $\I$ is the moment of inertia of the $SU(2)$ soliton determined from
the properties of the $N$ and $\Delta$.
As suggested by Manohar et al,$^{18}$
because of the $\delta$-function structure of
the meson wavefunctions and a parity-flip at the origin,
it is more convenient to transform the heavy-meson fields to
\be
\phi &\to& \phi' =  \phi \, \xi_0 , \nonumber \\
\phi^*_\mu &\to& \phi_\mu^{*'} = \phi_\mu^* \, \xi_0, \nonumber \\
\vec n &\to& \xi_0\,  \vec n \, \xi^\dagger_0.
\label{fieldredef}
\ee
The binding energy is not affected by this transformation.
With the primed fields, $\vec Q$ is of the form
\be
\vec Q &=& - \frac{1}{2} \int \! d^3 r
\left\{ \phi' \left( \Sigma^\dagger \vec \tau
\Sigma + \vec \tau \right) \phi'^\dagger + \phi^{*'}_i
\left( \Sigma^\dagger \vec \tau
\Sigma + \vec \tau \right) \phi^{*'\dagger}_i \right\}.
\label{Qprime}
\ee
Now since in the soliton rotating frame, the
``isospin" of the meson is transmuted to spin, we can identify
\be
\vec Q = c \vec J_Q,
\ee
namely, proportional to
the angular momentum lodged in the meson which is $1/2$.
Canonical quantization of (\ref{canonical}) leads to an
$O(m_\Phi^0 \cdot N_c^{-1})$ splitting\footnote{Modulo a hidden $m_\Phi^{-1}$
dependence in $c$ explained below.} in energy given
by the Hamiltonian
\be
\Delta H= 2 \I \Omega^2=\frac{1}{2\I} \left(\vec{J}_l +c\vec{J}_Q\right)^2
\label{analog}
\ee
with the spectrum
\be
\Delta E_{hf} = \frac{1}{2\I} \left\{ c J(J+1) + (1-c)J_\ell(J_\ell+1)
+ c(c-1) J_Q (J_Q+1) \label{hfsplit}
\right\},
\ee
where $\vec J_\ell$ is the spin lodged in the rotor.
The total spin $\vec J$ of the system is
\be
\vec{J}=\vec J_\ell + \vec J_Q.
\ee

The Hamiltonian (\ref{analog})
is a heavy-baryon analog to the diatomic molecular spectrum (\ref{theH})
and $c$ an analog to $(1-\kappa)$.
One can show by an explicit calculation that with (\ref{Qprime})
\be
c=0 .
\ee
This is the analog of the vanishing $(1-\kappa)$ in diatomic molecules,
a consequence of restored rotational symmetry. What happens here is that
the first term of (\ref{Qprime})
coming from the $P$ mesons gets cancelled exactly by the second
coming from the $P^*$ mesons. If the $P$ and $P^*$ were not degenerate
the cancellation would not occur. This suggests the following:
For not too large $m_\Phi$, say, $m_K$, $c$ can be substantial,
of $O(1)$, since the $K^*$ is rather high-lying compared with the
$K$. As $m_\Phi$ becomes large, the $P^*$ comes near the $P$, thus decreasing
$c$ such that in the heavy-quark limit, we get $c=0$.

Given that $c=0$ in the Isgur-Wise limit, we have the splitting
\be
\Delta E_{hf} = \frac{1}{2\I}  J_\ell ( J_\ell +1 ).
\label{energyp}
\ee
This $\Delta E_{hf}$ predicts
that there is an effective ``fine"
splitting of right sign and magnitude
between $\Lambda$ and the degenerate $\Sigma$ and $\Sigma^* $.
The predicted mass spectrum (denoting the mass by
the particle symbol) for b-quark baryons, with $\Lambda- N=4.65$ GeV
to fix $\alpha$, is
\be
{\Sigma_b} - N
= {\Sigma_b^*} - N &=& 4.84 \mbox{ GeV}.
\ee
These are comparable to the predictions of
quark potential models
\be
({\Lambda_b} - N)^{\mbox{\tiny QM}}
&=& 4.65 \mbox{ GeV}, \nonumber \\
({\Sigma_b} - N)^{\mbox{\tiny QM}}
&=& 4.86 \mbox{ GeV}, \nonumber
\ee
and to those of bag models
\be
({\Lambda_b} -  N)^{\mbox{\tiny BM}}
&=& 4.62 \mbox{ GeV}, \nonumber \\
({\Sigma_b} -  N)^{\mbox{\tiny BM}}
&=& 4.80 \mbox{ GeV}.\nonumber
\ee
\subsection{Hyperfine spectrum}

It is possible, within the scheme described so far, to discuss hyperfine
splitting with a nonzero $c$. For a finite heavy-quark mass for which
$m_\Phi < m_\Phi^*$,
the CK model indicates that $c \sim 1/m_\Phi$. This is the hidden $m_\Phi^{-1}$
dependence buried in the hyperfine coefficient $c$ alluded above
which we conjecture may have an intricate connection to a Berry potential.
For a sufficiently large $m_\Phi$, we may therefore
assume $c=a/m_\Phi$. Now using
(\ref{hfsplit}), we can write for baryons with
one heavy quark $Q$
\be
\Sigma_{\mbox{\tiny Q}}-\Lambda_{\mbox{\tiny Q}}=\frac{1}{\cal I}
(1-c_{\mbox{\tiny Q}})
\simeq 195 {\rm MeV} (1-c_{\mbox{\tiny Q}}). \label{finesplit}
\ee
With the experimental value $\Sigma_c-\Lambda_c\approx 168 {\rm MeV}$ for
the charmed baryons, we get $c_{c}\simeq 0.14$. This means that with
$m_{\mbox{\tiny D}}=1869 {\rm MeV}$, the constant $a\simeq 262$ MeV. So
\footnote{It is amusing to note that this formula works satisfactorily
even for the kaon for which one predicts $c_s \simeq 0.53$
to be compared with the empirical value 0.62.}
\be
c_{\Phi}\simeq 262 {\rm MeV}/m_\Phi.
\ee
Now for b-quark baryons, using $m_{\mbox{\tiny B}}=5279$ MeV, we find
$c_b\simeq 0.05$ which with (\ref{finesplit}) predicts
\be
\Sigma_b-\Lambda_b\approx 185 {\rm MeV}.
\ee
This agrees well with the quark-model prediction. Furthermore
the $\Sigma^*-\Sigma$
splitting comes out correctly also. For instance, it is predicted that
\be
\frac{\Sigma^*_b-\Sigma_b}{\Sigma^*_c-\Sigma_c}\simeq
\frac{m_{\mbox{\tiny D}}}{m_{\mbox{\tiny B}}}\approx 0.35
\ee
to be compared with the quark-model prediction $\sim 0.32$.
If one assumes that the heavy mesons $\Phi$ are weakly interacting, then
we can put more than one $\Phi$'s in the soliton $^{10}$ and obtain the spectra
for $\Xi$'s and $\Omega$'s in good agreement with quark-model results.$^{19}$

\section{Discussions}

I have discussed how one can continuously ``dial" from chiral symmetry
to IW symmetry in the spectrum of baryons. This is a surprising outcome.
In doing so, a Wess-Zumino
like term plays an essential role. For $m_Q < \Lambda_\chi$, the Wess-Zumino
term plays a key role in binding a pseudoscalar $\Phi$ to an $SU(2)$
soliton. For $m_Q >>\Lambda_\chi$, the Wess-Zumino term vanishes but
a term of the form ${\rm Tr}\bar{H}v_\mu H B^\mu$ contributes through
coupling with the light vector meson $\omega$ which given a reasonable
strength
again dominates the binding. The baryon structure is exactly the same
as the one given by the CK model with the meson shrunk to the center
of the soliton. It is not clear whether all this is just a coincidence
or something profound but it is certainly intriguing.

The description of Manohar and his collaborators $^{18}$ differs from
the above scenario in that in their scheme,
the heavy meson $H$ gets bound to a {\it
rotating} skyrmion through a residual interaction given by ${\L}_P$
in (\ref{Ham}) with no contribution from a Wess-Zumino like term.
Since one starts here with a Lagrangian in the IW symmetry limit, the hyperfine
splitting comes from an IW symmetry breaking term of the form
\be
\frac{c}{m_H} {\rm Tr} (\bar{H}\sigma_{\mu\nu}H\sigma^{\mu\nu})
\ee
which splits the degeneracy of $P$ and $P^*$.
Surprisingly the results of both approaches seem rather close. The connection
between the two is not yet understood.

In terms of the ``Berry charge," the limit $c=0$ clearly corresponds to
the restoration of the IW symmetry, namely the symmetry that makes
$P$ and $P^*$ degenerate. It remains to be seen how this result can be
obtained in the general setting of Berry potentials in the strong
interactions
developed recently by Lee, Nowak, Zahed and myself.$^{20}$ More work is
needed in this area.

A matter of potential interest on which I have no result to discuss is
the possibility of having the chiral partner $S$ of $H$ which
arises naturally in approximate bosonization of QCD $^{15}$ bound to
a soliton. A rich variety of spectra associated with this excitation
is predicted.

\nonumsection{Acknowledgments}

I would like to acknowledge fruitful discussions with Hyun Kyu Lee,
Dong-Pil Min, Maciej Nowak, Yongseok Oh, Dan Olof Riska, Norberto Scoccola and
Ismail Zahed.

\nonumsection{References}

\end{document}